\documentclass[rnote]{aa} % for the research notes
\usepackage{natbib}
\usepackage{graphicx}
\usepackage{txfonts}
\begin{document}

   \title{Four new HgMn stars: HD 18104, HD 30085, HD 32867, HD 53588}

   \subtitle{}

%----------------------
  \author{R. Monier\inst{1,2}, M. Gebran\inst{3}, and F. Royer\inst{4}}

\institute{$^1$LESIA, UMR 8109, Observatoire de Paris, place J. Janssen, Meudon\\
\email{Richard.Monier@obspm.fr}\\  
$^2$Laboratoire Lagange, Universit\'e de Nice Sophia, Parc Valrose,            06100 Nice, France\\
\email{Richard.Monier@unice.fr}\\
$^3$ Department of Physics and Astronomy, Notre Dame University-Louaize, PO Box 72, Zouk Mikael, Lebanon\\
\email{mgebran@ndu.lb.edu}\\
$^4$ GEPI, Observatoire de Paris, place J. Janssen, Meudon, France\\
\email{Frederic.Royer@obspm.fr}}
%--------------------------------
\date{Received ; accepted}

\authorrunning{R. Monier et al.}
\titlerunning{Four new HgMn stars}
% \abstract{}{}{}{}{} 
% 5 {} token are mandatory
 
  \abstract
   {We have detected four new HgMn stars, while monitoring a sample of apparently slowly rotating superficially normal bright late B and early A stars in the northern hemisphere.}
   {\textbf{Important classification lines of \ion{Hg}{ii} and \ion{Mn}{ii} are found as conspicuous features  in the high resolution SOPHIE spectra of these stars ($R=75000$)}.}
   {Several lines of \ion{Hg}{ii}, \ion{Mn}{ii} and \ion{Fe}{ii} have been synthesized using model atmospheres and the spectrum synthesis code SYNSPEC48 including  hyperfine structure of various isotopes when relevant. These synthetic spectra have been  compared to high resolution high signal-to-noise observations of these stars in order to derive abundances of these  key elements.}
   {The four stars are found to have distinct enhancements of Hg and Mn which show that these stars are not superficially normal B and A stars, but actually are new HgMn  stars and should reclassified as such.}
   {}


   \maketitle
%
%________________________________________________________________

\section{Introduction}
   
   We have recently undertaken a spectroscopic survey of all apparently slowly rotating bright early A stars (A0-A1V) and late B stars (B8-B9V) observable from the northern hemisphere. The incentive is 
to search for rapid rotators seen pole-on or new chemically peculiar B and A stars which have thus far remained unnoticed.
   The abondance results for the A0-A1V sample have been published in \cite{Royer}.
   The selection criteria were: a declination higher than $-15$\degr, spectral class A0 or A1 and luminosity class V and IV and magnitudes $V$ brighter than 6.65. The B8-9 sample employs the same criteria (\textbf{except for the V magnitude brighter than 7.85 as these B stars are intrinsically brighter in the V band where SOPHIE reaches its maximum efficiency}. Most of the stars of that B8-9 sample (40 stars) have just been observed in December 2014.
   A careful abundance analysis of the high resolution high signal-to-noise ratio spectra of the A stars sample has allowed to sort out the sample of 47 A stars into 17 chemically normal stars (whose abundances do not depart by more  than $\pm 0.20$ dex from solar values), 12 spectroscopic binaries and 13 chemically peculiar stars (CPs) among which five are new   CP stars. The status of these new CP stars still needs to be fully specified by spectropolarimetric observations to address their magnetic nature or by exploring new spectral ranges which we had not explored in this first study. Indeed, the abundance analysis of the A stars sample in \cite{Royer} relied only on four spectral regions: 4150--4300\,\AA, 4400--4790\,\AA, 4920--5850\,\AA, and 6000--6275\,\AA, avoiding Balmer lines and atmospheric telluric lines.
   \\
   We have now started to examine the A0-A1V and B9-B8V samples in the region of the red wing of the H${\epsilon}$ line to seach for the  \ion{Hg}{ii} 3983.93\,\AA\ line and also the spectral range 4125--4145\,\AA\ which harbors  the \ion{Mn}{ii} line at 4136.92\,\AA\ next to the \ion{Si}{ii} doublet  (Multiplet 2). The presence of the
\ion{Hg}{ii} 3983.93 \,\AA\ line and   the \ion{Mn}{ii} line at 4136.92\,\AA\ lines  
 with appreciable strengths are the signatures of an HgMn star \citep{Jaschek}.
 \textbf{Note that, a priori, the presence of only one of these two elements (Mn or Hg) is actually sufficient to identify a new HgMn star as both elements may be distributed in different spots and may not be simultaneously visible to the observer at a given time (see Hubrig et a. (2012) for a complete review).}
 In our A and B stars samples, we found that four stars definitely show the \ion{Hg}{ii} line at 3984\,\AA\ and the \ion{Mn}{ii} line at 4136.92\,\AA\ (and several other strong \ion{Mn}{ii} lines). These are  HD 18104 (B9V), HD 30085 (A0IV), HD 32867 (B8V), HD 53588 (B9V). A bibliographic query of the CDS for each of these stars actually reveals very few publications (about 10 references each). HD 30085 was ascribed a spectral type A0IV in \cite{Cowley1969} survey of bright A stars (observed with prismatic dispersion of 125\,\AA\,mm$^{-1}$ around H${\gamma}$)
 and the authors did not mention any peculiarity in the spectrum. The other three B stars do not appear in \citeauthor{Cowley1972}'s classification (\citeyear{Cowley1972}) of the bright B8 stars.
The purpose of this research note is to report on the detection of the \ion{Hg}{ii} 3983.93\,\AA\ line and strong \ion{Mn}{ii} lines as well in these four stars classified currently as ``normal''. We also have determined estimates of the iron, manganese and mercury abundances using spectrum synthesis to quantify the enhancements of these elements in these stars.

 \section{Observations and reduction}
 The four stars have been observed at Observatoire de Haute Provence using the high resolution ($R = 75000$) mode of the SOPHIE echelle spectrograph \citep{Perruchot} at three different epochs: February 2012, December 2013 and December 2014. 
 The three late B stars have been observed only once in December 2014. The A0IV star HD 30085 was observed twice, in February 2012 and December 2013. The coaddition of the two exposures for HD 30085 yields a coadded spectrum whose $\frac{S}{N}$ is about 300. The three late B stars observed in December 2014 have lower $\frac{S}{N}$ ratios ranging from 136 to 170. The observations log is displayed in Table\,\ref{table:1}.
 The data are automatically reduced to produce 1D extracted and wavelength calibrated \'echelle orders. Each reduced order was normalised separately using a Chebychev polynomial fit with sigma clipping, rejecting points above or below 1 $\sigma$ of the local continuum. Normalized orders were merged together, corrected  by the blaze function and resampled into a constant wavelength step of about 0.02\,\AA\ \cite[see][for more details]{Royer}.
 
Radial velocities were derived from cross-correlation techniques, avoiding the Balmer lines and the atmospheric telluric lines. The normalized spectra were cross-correlated with a synthetic template extracted from the POLLUX database\footnote{http://pollux.graal.univ-montp2.fr}  \citep{Pas_10} corresponding to the parameters $T_\mathrm{eff}=11000$\,K, $\log g=4$ and solar metallicity. A parabolic fit of the upper part of the resulting cross-correlation function gives the Doppler shifts, which is then used to shift spectra to rest wavelengths. \textbf{The radial velocity of HD 30085 is found to be the same within the accuracy at the two different epochs of observations:
$V_{rad}$ = 8.27 $\pm$ 0.20 $km.s^{-1}$. We looked for line profile variability in the lines of Hg II and Mn II only}. 
\textbf{The individual spectra of HD 30085 were coadded and the difference of each spectrum with the mean spectrum computed. This difference was then ratioed to the estimated noise level $\sigma$ of the local continuum. We considered real variations only if the absolute value of the difference was found to be larger than 3 $\sigma$. We failed to find any real line profile variations, nor radial velocity variations above three sigmas of the noise level for HD 30085}.

 \section{The Hg II and Mn II lines in the four stars}
 
 Two spectral regions have been used to establish the HgMn  nature of the four stars. First, the red wing of H${\epsilon}$,which lies in order 3, harbors the \ion{Hg}{ii} $\lambda$ 3983.93\,\AA\ line and several \ion{Zr}{ii} and \ion{Y}{ii} lines likely to be strengthened in HgMn stars. 
\cite{Jaschek} emphasize that the \ion{Hg}{ii} 3983.93\,\AA\ line is absent in the spectra of normal late B stars.
\textbf{Second, the region from 4125\,\AA\ to 4145\,\AA\  (order 6) contains the the  \ion{Mn}{ii} line at 4136.92\,\AA\ redwards of the \ion{Si}{ii} doublet. (Multiplet 2).
In an HgMn star, the \ion{Mn}{ii} line at 4136.92\,\AA\ should be strong wheras it should be absent in any comparison normal late B-type star.}
Furthermore, the  lines of \ion{Mn}{ii} at 4206.37\,\AA\ and 4252.96\,\AA\ should be enhanced too in the spectra of HgMn stars \citep{Gray}. 
 
\begin{figure}
   \vskip 0.75cm
   \centering
   \includegraphics[scale=0.35]{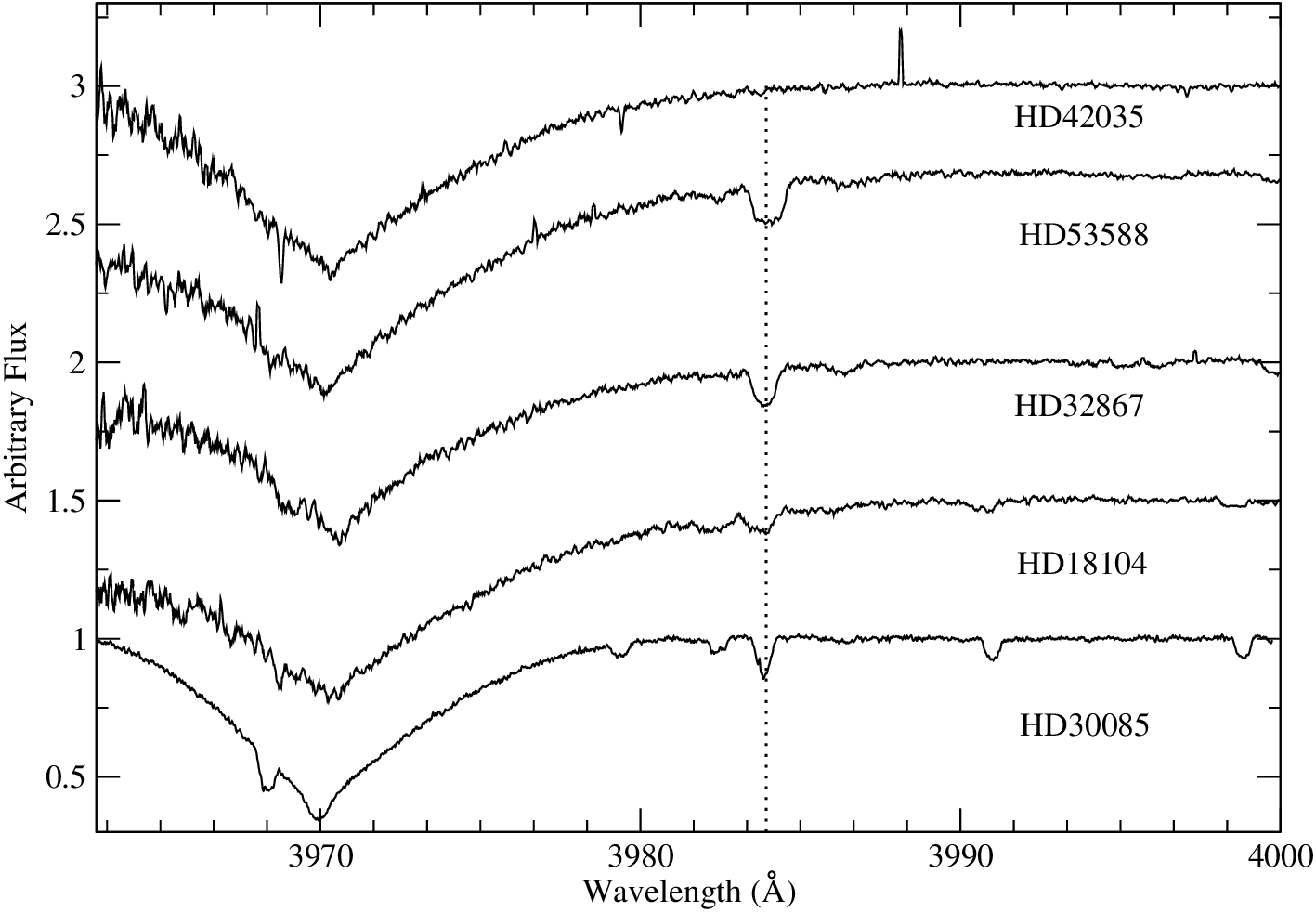}
      \caption{The detection of the \ion{Hg}{ii} line at 3984\,\AA\ , its rest wavelength location is depicted as a vertical line in the four stars. The spectra have been offset from each other for clarity. The \ion{Hg}{ii} line is absent in the spectrum of the comparison star HD 42035.}
         \label{HgIiAll}
   \end{figure}

\begin{figure}
     \vskip 0.75cm
 \centering
   \includegraphics[scale=0.35]{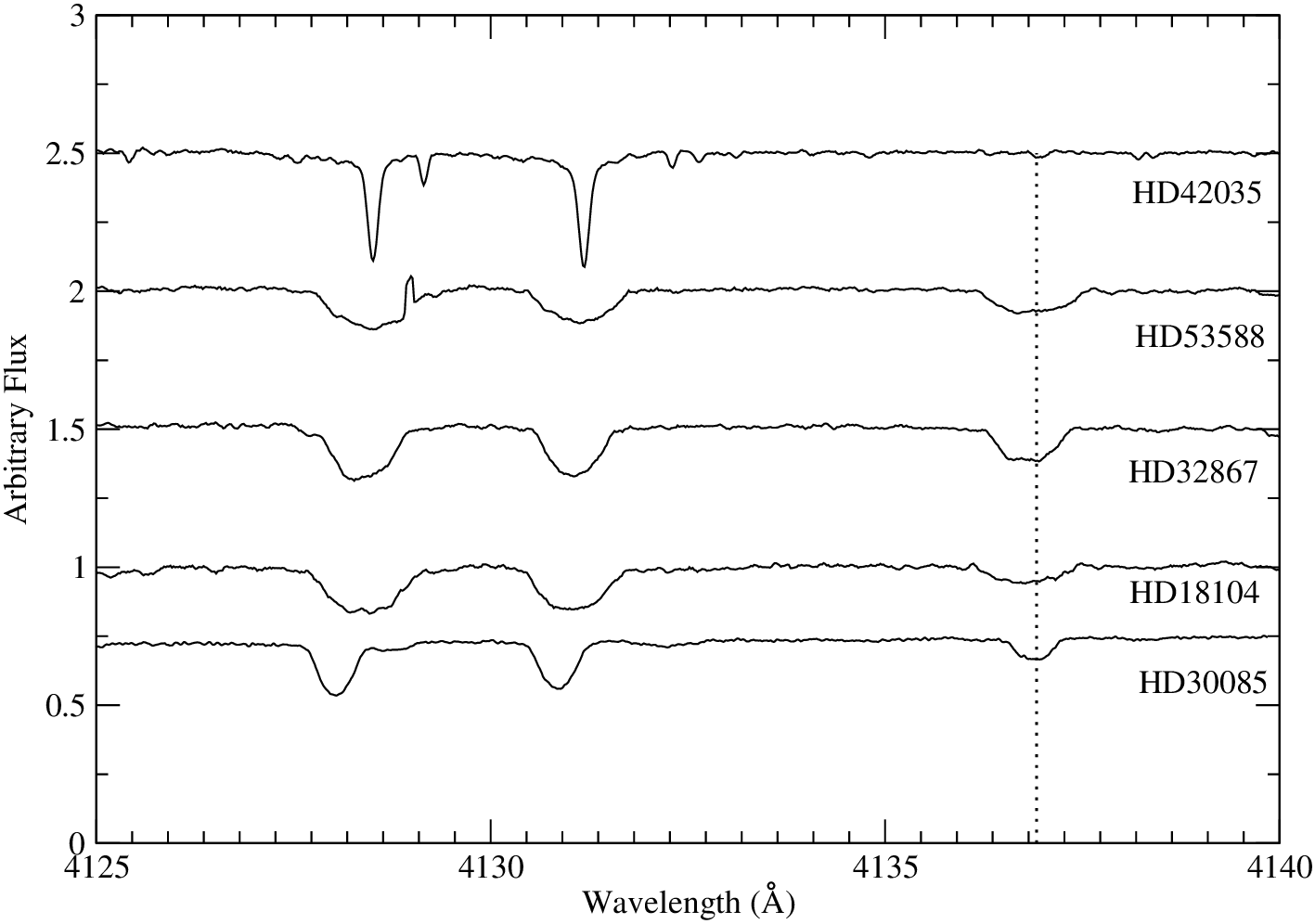}
      \caption{The detection of the Mn line at 4136.92\,\AA\ , its rest wavelength location is depicted as a vertical line in the four stars. The spectra have been offset from each other for clarity. The \ion{Mn}{ii} line is absent in the spectrum of the comparison star HD 42035.}
         \label{MnIIAll}
   \end{figure}

Figure\,\ref{HgIiAll} displays order 3 for the four stars and a comparison star, HD 42035 (a normal B9 V star of our sample), and the location as a vertical line of the rest wavelength of the  \ion{Hg}{ii} 3983.93\,\AA\ line (Multiplet 2), Fig.\,\ref{MnIIAll} displays the \ion{Mn}{ii} line at 4136.92\,\AA\ longwards of
the \ion{Si}{ii} doublet (Multiplet 2) at 4128.07\,\AA\ and 4130.88\,\AA. The \ion{Hg}{ii} 3983.93\,\AA\ line and the \ion{Mn}{ii} 4136.92\,\AA\ are clearly present in the four spectra of these stars and absent in the comparison star HD 42035. The equivalent width of the \ion{Hg}{ii} 3983.84\,\AA\ line ranges from 64 m\,\AA\ to 132 m\AA\  which is well in the range quoted  for HgMn stars \cite[from 50 m\AA\ up to 300 m\AA\ typically in][]{Jaschek}.  \textbf{The equivalent width (EW) of the \ion{Mn}{ii} line at 4136.92\,\AA\ ranges from 47.4 m\AA\ up to 93.6 m\AA\ in the four stars whereas this Mn II line is absent in the spectrum of the comparison star HD 42035.} The \ion{Mn}{ii} lines at 4206.37\,\AA\ and 4252.96\,\AA\  are also prominent features in the spectra of all stars. The EWs of 4206.37\,\AA\ ranges from 66.7 m\AA\ to 117.4 m\AA\  in agreement with \cite{Kodaira} who find a mean EW of about 100 m\AA\ for the 4206.37\,\AA\ line for a group of HgMn stars. The EW of 4252.96\,\AA\ ranges from 75 m\AA\ up to 122.4 m\AA. 

% Suppress IUE paragraph
%HD 32867 has 2  IUE spectra which we retrieved from the MAST database. SWP 276562 is a low resolution spectrum where it is difficult to assess the presence of the strongest expected  lines of Hg I (resonance line 1849.508\,\AA) and of \ion{Hg}{ii} (the 2 resonance lines at 1649.947 and 1942.287\,\AA).
 % figure 1: \ion{Hg}{ii} region order 3
 
 % figure 2: \ion{Mn}{ii} region order 6
 
% Table 11 version Marwan

 \begin{table}
\caption{Observation log} % title of Table
\label{table:1} % is used to refer this table in the text
\centering % used for centering table
\begin{tabular}{c c c c cc} % centered columns (5 columns)
\hline\hline % inserts double horizontal lines
Star ID & Spectral & V &Observation & Exposure & S/N \\ % table heading
& type&  & Date & time (s ) &\\ % table heading
\hline % inserts single horizontal line
HD 18104 & B9    & 6.85 & 2014-12-16 & 800 & 136\\ % inserting body of the table
HD 30085 & A0IV & 6.35 & 2013-12-11 & 1200 & 269 \\
                &         &        & 2012-13-03   & 800   &  216 \\
HD 32867 & B8V & 7.48  & 2014-12-16  & 2100 & 170\\
HD 53588 & B9V & 7.20  & 2014-12-17  & 1350 & 158 \\
\hline %inserts single line
\end{tabular}
\end{table}
%_____________________________________________________________
%                Figure order 3 \ion{Hg}{ii} line
%-------------------------------------------------------------
   
 %-------------------- Figure order 6  \ion{Mn}{ii} line -------------------------------------------  

 \section{Abundance determinations}
 
 \subsection{Fundamental parameters determinations}
 
 For the three B stars, we have adopted the effective temperatures and surface gravities derived by \cite{Huang} from fitting the H${\gamma}$ profiles.
 %Note that the effective temperature and surface gravity for HD 32867 can also %be derived using Napiwotzky's (\citeyear{Napiwotzki}) UVBYBETA procedure
% and respectively  yield $T_\mathrm{eff}$ = 14074 K and $\log g = 3.92$ which %only slightly differ from the values derived in \cite{Huang}.  
\textbf{Indeed two B stars, HD 18104 and HD 53588, do not have Str\"{o}mgren's photometry which precludes using Napiwotzky's (\citeyear{Napiwotzki}) UVBYBETA procedure to derive their fundamental parameters}.
We therefore used the effective temperature
 and surface gravity derived in \cite{Huang} by fitting the H${\gamma}$ profile for the three stars for consistency. For HD 30085, the effective temperature and surface gravity were derived from Str\"{o}mgren's photometry in \cite{Royer} and a spectrum synthesis of its H${\gamma}$ profile was run to confirm these parameters. The adopted effective temperatures, surface gravities and projected equatorial velocities are displayed in Table\,\ref{table:2}.   
 
 \begin{table}
\caption{Stellar fundamental parameters}             % title of Table
\label{table:2}      % is used to refer this table in the text
\centering                          % used for centering table
\begin{tabular}{c c c c c}        % centered columns (4 columns)
\hline\hline                 % inserts double horizontal lines
Star ID & $T_\mathrm{eff}$ & $\log g$  & $v\sin i$ & $V_{rad}$\\    % table heading 
            &               &               &  (km\,s$^{-1}$) & (km\,s$^{-1}$)\\
\hline                        % inserts single horizontal line
   HD 18104 & 11074 & 3.67 & 46.0 & 12.07 \\      % inserting body of the table
   HD 30085 & 11300  & 3.95  & 23.0 &   8.20  \\
   HD 32867 & 13149 & 3.86  & 37.0  & 13.97 \\ 
   HD 53588 & 12351 & 3.88 &  48.0  & 12.03 \\
\hline                                   %inserts single line
\end{tabular}
\end{table}

 \subsection{Model atmospheres and spectrum synthesis calculations}
 
  Plane parallel model atmospheres assuming radiative equilibrium and hydrostatic equilibrium were computed using the ATLAS9 code \citep{Kurucz}. The linelist was built from \cite{Kurucz} gfhyperall.dat\footnote{http://kurucz.harvard.edu} which includes hyperfine splitting levels. A grid of synthetic spectra was computed with SYNSPEC48 \citep{Hubeny} to model the \ion{Hg}{ii}, \ion{Mn}{ii}, \ion{Fe}{ii} and \ion{Si}{ii} lines. Computations were iterated varying the unknown abundance until minimization of the chi-square between the observed and synthetic spectrum was achieved. The microturbulent velocities have been assumed to be  0 or 0.5 km\,s$^{-1}$ in agreement with most analyses of HgMn stars whose atmospheres are thought to be very quiet.
 
 \begin{table}
\caption{Abundance determinations} % title of Table
\label{table:3} % is used to refer this table in the text
\centering % used for centering table
\begin{tabular}{c c c c } % centered columns (4 columns)
\hline\hline % inserts double horizontal lines
Star ID & Chemical & Laboratory & $[\frac{N_{elem}}{N_{solar}}]$  \\ % table heading
& Element & wavelength (\AA) &  $\odot$ \\ % table heading
\hline % inserts single horizontal line
HD 18104 & \ion{Fe}{ii} & 4500-4550   &   2.5  $\pm$ 0.5\\ % inserting body of the table
HD 30085 & \ion{Fe}{ii}  & 4500 - 4550 &   2.5 $\pm$ 0.5  \\
HD 32867 & \ion{Fe}{ii}  & 4500 - 4550  &   2.0  $\pm$ 0.5 \\
HD 53588 & \ion{Fe}{ii}  & 4500 - 4550  &   2.5  $\pm$ 0.5 \\
\hline
HD 18104 & \ion{Mn}{ii} &  4136.92  &    50  $\pm$ 4.5 \\ % inserting body of the table
HD 30085 & \ion{Mn}{ii} &  4136.92  &   40 $\pm$  4.0\\
HD 32867 & \ion{Mn}{ii}  & 4136.92  &  350 $\pm$ 32.0   \\
HD 53588 & \ion{Mn}{ii}  & 4136.92 &  150 $\pm$   14.0  \\
\hline
HD 18104 & \ion{Hg}{ii} &  3983.87  & 40000 $\pm$  2800\\ % inserting body of the table
HD 30085 & \ion{Hg}{ii}  &  3983.87  &   32000   $\pm$ 2000 \\
HD 32867 & \ion{Hg}{ii} & 3983.87  &  130000 $\pm$  9100 \\
HD 53588 & \ion{Hg}{ii}  & 3983.87  &  300000  $\pm$  27000 \\
\hline %inserts single line
\end{tabular}
\end{table}

\subsection{The derived iron, manganese and mercury abundances}

\begin{figure}
      \vskip 0.75cm
   \centering
      \includegraphics[scale=0.35]{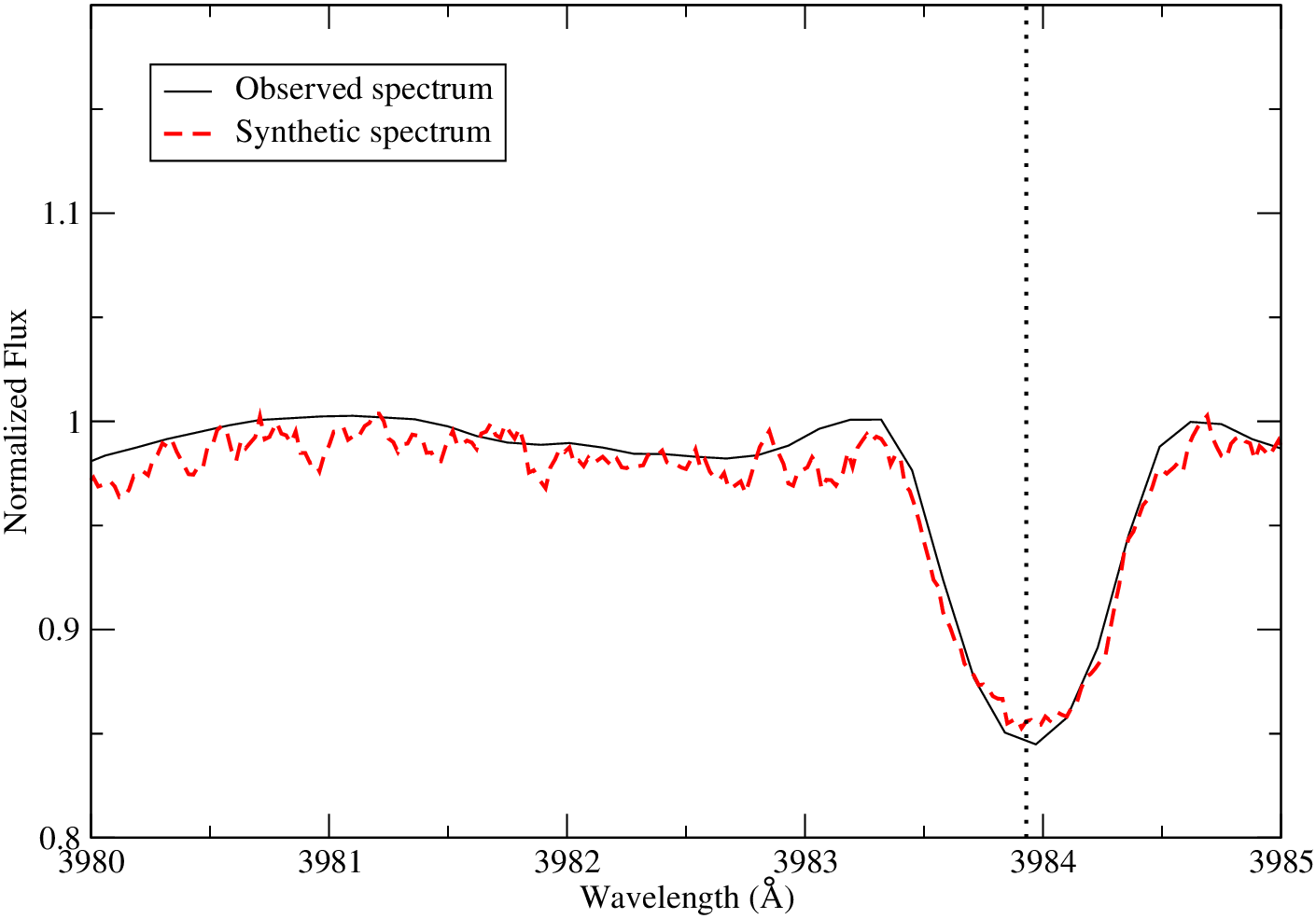}
   \caption{Comparison of the observed \ion{Hg}{ii} 3983.93\,\AA\ profile of HD 32867 to a synthetic profile computed for an overabundance of 130000 solar. The vertical line depicts the location of the rest wavelength of the \ion{Hg}{ii} line}
   \label{HgIIHD32867}
 \end{figure}

The iron abundances have been derived by using several \ion{Fe}{ii}  lines of multiplets 37, 38 and 186  in the range 4500--4600\,\AA\ whose atomic parameters are critically assessed in NIST\footnote{http://www.nist.gov} (these are C+ and D quality lines). These lines are widely spaced and the continuum is fairly easy to trace in this spectral region. Their synthesis always yields consistent iron abundances from the various transitions with very little dispersion. The iron abundance is probably the most accurately determined of the three abundances derived here.
We find that the four stars show only mild enhancement in iron, about 2 to 2.5 solar.  The found \ion{Fe}{ii}, \ion{Mn}{ii}, and \ion{Hg}{ii} abundances and their estimated uncertainties for the four selected stars are collected in Table\,\ref{table:3} \cite[the determination of the uncertainties is discussed in][]{Royer}.
\\
The manganese abundances have been derived from two C+ quality lines, $\lambda$ 4206.368\,\AA\ and 4259.191\,\AA\ which have hyperfine structure published in \cite{Holt}. The individual transitions were actually added to our intitial lineslist using the wavelengths, oscillator strengths and angular momenta from Table 1 in \cite{Holt}. We have not used the other two \ion{Mn}{ii} lines at 4326.644\,\AA\ and 4348.396\,\AA\ analysed by \cite{Holt} as they fall in the blue and red wings of the H${\gamma}$ line where the continuum is more difficult to locate. Once the hyperfine structure of 4206.368\,\AA\ and 4259.191\,\AA\ are properly taken into account, these two \ion{Mn}{ii} lines yield very consistent abundances which are significantly lower than when hyperfine structure is ignored (up to a factor of 10 lower). We find manganese enrichments ranging from 40 up to 350 solar, which appear to slowly increase with effective temperature, the two coolest stars (HD 18104 and HD 30085) around 11000K having the lowest manganese enrichment (40--50 $\odot$)  while the hottest star, HD 32867, is the most enriched. This agrees with the correlation of manganese abundances with the effective temperature reported by \cite{Smith} using ultraviolet lines of \ion{Mn}{ii} for a large number of HgMn stars.
\\
The mercury abundances have been derived from the \ion{Hg}{ii} 3983.93\,\AA\ line in NIST (multiplet 2) but including the hyperfine structure of the various isotopes as provided in \cite{Dolk}. We have used the wavelengths and oscillator strengths for each transition as presented in their Table 2 for the case of a terrestrial isotopic mixture. The other line of \ion{Hg}{ii} that might be present in the spectra of HgMn stars is that of multiplet 4 at 6149.48\,\AA\ but it appears to be blended in all four stars with the \ion{Fe}{ii} line at 6149.25\,\AA\ so that we did not use this line for abundance analysis. \cite{Dolk} assume a terrestrial isotope mixture for mercury which may not occur in the HgMn stars we are looking at.
Note that the 3983.93\,\AA\ line is actually blended with two lines, \ion{Cr}{i} 3983.897\,\AA\ and \ion{Fe}{i} 3983.960\,\AA, which contribute no absorption at the temperatures of the late B stars we study here (this is easily verified by removing the \ion{Hg}{ii} hyperfine transitions from the line synthesis and by checking that no absorption is computed for the appropriate iron abundance for none of these stars).
The found Hg overabundances range from 32000 to 300000 solar where again the coolest stars tend to have lower Hg enhancement than the hotter ones. The uncertainties on these abundances are believed to be of the order of $\pm 2000$ $\odot$ to $\pm 10000$ $\odot$.
The synthetic spectrum reproducing the best the \ion{Hg}{ii} profile for HD 32867 for a 130000 solar overabundance of Hg is compared to the observed profile rectified to the red wing of H${\epsilon}$ in Fig.\,\ref{HgIIHD32867}.
Only the mean spectrum of HD 30085 is of sufficient quality to examine the structure of the line core. In the two observations of this star, the core of the 3983.93\,\AA\ line appears to be fairly flat and extending from 3893.90 $\pm 0.02$\,\AA\ to 3984.07 $\pm 0.02$\,\AA, which roughly correspond to the positions of the transitions of the heaviest isotopes \element[][200]{Hg}  and \element[][204]{Hg}.
The accuracy of the wavelength scale is achieved by using four control lines on each side of the \ion{Hg}{ii} line: shortwards the \ion{Zr}{ii} line at 3982.025\,\AA, the \ion{Y}{ii} line (M 6) at 3982.59\,\AA\ and longwards the \ion{Zr}{ii} lines at 3984.718\,\AA\ and 3991.15\,\AA: after correction for the radial velocity of HD30085, the centers of these lines are found at their expected laboratory locations to within $\pm 0.02$\,\AA). These two isotopes appear thus to contribute most of the absorption in HD 30085 as is the case in the majority of the coolest HgMn stars \citep{White}, whereas they contribute only a small fraction of the terrestrial mixture. 
% Note that the \element[][204]{Hg} isotope is a pure r process product whereas  % \element[][200]{Hg} can be produced by either the r process or the s process. 

% Figure 3 

\section{Conclusions}

We have found four new HgMn stars from the inspection of the H${\epsilon}$ wing and a few \ion{Mn}{ii} lines. We provide for the first time estimates of the  overabundances of Fe, Mn and Hg in these stars, until now considered as normal. We believe that they must be reclassified as HgMn stars. The detection of these new HgMn  stars is quite important as they only represent about 8\% for the coolest B-type stars around B9 and B8 \citep{Wolff}. We expect to find other new HgMn stars by extending our survey to the late B stars of the  southern  hemisphere. Each of these stars will be the subject of a detailed abundance analysis which we plan to publish in the near future.

\begin{acknowledgements}
We thank the referee, Dr. S. Hubrig, whose comments resulted in many improvements. We also thank the OHP night assistants for their helpful support during the three observing runs.
\end{acknowledgements}

%-------------------------------------------------------------------

\end{document}